\documentclass[aps,epsf,preprint,nofootinbib]{revtex4}

\def\1{{\bf 1}}
\def\[{\left[}
\def\]{\right]}
\def\be{\begin{eqnarray}}
\def\ee{\end{eqnarray}}
\def\bn{\begin{enumerate}}
\def\i{\item}
\def\en{\end{enumerate}}
\def\nn{\nonumber}
\def\({\left(}
\def\){\right)}
\def\bk#1{\langle#1\rangle}
\def\eq#1{(\ref{#1})}

\def\o{\omega}

\def\f{\phi}

\def\G{{\cal G}}
\def\R{{\cal R}}

\def\l{\lambda}

\def\r2{\sqrt{2}}

\def\da{{\bf 1}}
\def\dc{{\bf 3}}
\def\db{{\bf 2}}

\def\diag{{\rm diag}}
\def\x{$\times$}
\def\bl{$\bullet$}

\begin{document}

\title{A bottom-up analysis of horizontal symmetry}
\author{C.S. Lam}
\address{Department of Physics, McGill University\\
 Montreal, Q.C., Canada H3A 2T8\\
and\\
Department of Physics and Astronomy, University of British Columbia,  Vancouver, BC, Canada V6T 1Z1 \\
Email: Lam@physics.mcgill.ca}

\begin{abstract}
The group-theoretical method used to derive horizontal symmetry from neutrino mixing is reviewed and expanded.
Some misunderstanding in the literature regarding the result is clarified. 
The method used previously to find vacuum alignments of $S_4$
is applied to compute those of $A_4$ and $S_3$. A study of 
effective theories based on these three groups 
shows that in each case there are just enough free parameters
to fit all the masses and the remaining mixing parameters. 
This places constraint on dynamical models because effective theories are 
just dynamical models with the right-handed fermions integrated out.
 How quarks may fit into this  scheme is briefly discussed.
\end{abstract}
\narrowtext
\maketitle

\section{Introduction} 
The usual method to model a mass matrix proceeds from the top down. 
 First, a horizontal symmetry group $\G$ is picked, an irreducible
representation (IR) is assigned to each left-handed (LH) and each right-handed (RH) fermion 
in the theory, and a number of Higgs-like scalar bosons 
is created to couple 
invariantly to the fermions. Then, Higgs expectation values are 
introduced to break the symmetry, from which mass matrices are computed. 
The mass matrices so obtained depend on the Yukawa couplings, the mass parameters, and the Higgs expectation values. 
These parameters are chosen to fit the observed fermion masses, the mixing angles, and the phases. Such a construction is often applied to the leptons alone, 
but sometimes it is also carried out in
grand unified theories containing both leptons and quarks. In the latter case frequently only approximate fits
are obtained.

A model containing an equal or a larger number of tunable parameters than available experimental quantities
 can be expected to fit all the data, even without a 
horizontal symmetry. Thus the introduction of a horizontal symmetry is justified only
when it leads to unambiguous predictions to allow a smaller number of parameters 
 to be used to fit the data. Many existing models do have fewer parameters, some much fewer,
but if they fit the experimental quantities only approximately, then it is unclear what and where 
exactly are the savings coming from the symmetry assumption.

In this article we review and expand on a complementary approach \cite{LAM}, 
{\it using only left-handed
fermions} and {\it start from the bottom up}. 
In this approach, symmetry is directly connected to mixing, so the gain from  symmetry is clear.
The simplification of using only LH fermions comes about 
 because low energy data such as masses and mixing parameters can all be
determined from the effective mass matrices connecting only LH fermions.

The simplification also brings about its limitation, in that 
without the RH
fermions local dynamical theories cannot be constructed. 
In other words, although the connection between mixing and the LH effective theory is straight forward and unique,
the next step of constructing a dynamical model is not.
Nevertheless, this analysis tells us what would happen
to every dynamical model after its RH components are integrated out. To the extent that low energy data can all be determined from the
LH effective theories, they already carry most of the information reachable by present-day experiments. 

Sec.~2 contains a review of the general method and the previous results. This method is used to compute the 
vacuum alignments of $A_4$ and $S_3$ in Sec.~3.  
The results are listed in
Table 1, together with the alignments of $S_4$ previously obtained and the IR from which 
these vacuum alignments are calculated.

A detailed discussion of the LH effective theories for these groups is carried out in Sec.~4. The number of 
independent parameters is determined in each case, with the result tabulated in Table 2. 
It turns out that in all the cases considered, 
there are always just enough parameters to fit all the masses and the undetermined
mixing parameters.

In principle, this bottom-up group-theoretical approach can be applied
to the quark sector as well, but the symmetry so obtained is very different from the symmetry in the lepton sector, 
because their mixing matrices are vastly different. For reasons already discussed \cite{LAM}, we prefer to determine the horizontal symmetry
from the lepton sector. The question of how quarks can  fit into this scheme will be briefly discussed in Sec.~5.

The remainder of this section is devoted to addressing some confusions found in the literature regarding our
previous results \cite{LAM}, summarized below.
\bn
\i If the neutrino mixing matrix is tri-bimaximal (TBM) \cite{TBM}, then the minimal left-handed (LH) symmetry group
\footnote{This is the horizontal symmetry group for an effective theory in which only LH fermions
and composite Higgs appear. Such an effective theory is obtained from a dynamical model by integrating out 
its RH fermions. The resulting effective Hamiltonian is generally non-local, but that does not
matter because we are only interested in its horizontal-symmetry structure. This group was referred to as the 
natural symmetry group in \cite{LAM}, 
but since there is objection to that name \cite{AFM}, I will use
 a more descriptive terminology here and call it the LH symmetry group.}
in the lepton sector is $S_4$; `minimal' means that every other LH symmetry group contains
$S_4$ as a subgroup.
\i If only trimaximal mixing taken from the second column of the TBM is assumed in the mixing, 
then the smallest LH symmetry
group is $A_4$. It is not known whether this group is minimal or not.
\i If only bimaximal mixing taken from the third column of the TBM is assumed, then the smallest LH symmetry
group is $S_3$. It is not known whether this group is minimal or not.
\i If only the first column of the TBM is assumed, then the minimal symmetry group is again $S_4$. In principle
the mixing matrix so obtained need not be TBM, because neither trimaximal mixing nor bimaximal mixing is explicitly assumed.
\en

Starting from a symmetry group $\G$ so derived, one can compute the Higgs alignment needed
to regain the original mixing. This has been carried out in $S_4$ \cite{LAM}, and it will also be done
for $A_4$ and $S_3$ in Sec.~3. 

Obviously, additional soft breaking or some other vacuum alignment can be imposed on $\G$ so that
the original mixing input is 
not recovered. Conversely, one may fit all the experimental data including mixing
without even any symmetry as long as a sufficient number of parameters are provided. 
The present approach simply tells us that certain predictions on mixing can be obtained if an appropriate
LH group $\G$ and the appropriate vacuum alignments are used, and that this correspondence is one to one.

If $\G$ is the LH symmetry group, one can always construct a local dynamical model with $\G$ as the dynamical
symmetry, giving rise to the same prediction on mixing,
{\it but there is no claim to the converse}. In other words, it is conceivable that a dynamical model with
some other symmetry group can lead to a LH effective theory with an enhanced LH symmetry group $\G$.
With these remarks, one finds that
the two `counter examples' in \cite{GLL} really do not contradict the
results cited in  1 to 4 above.

\section{The connection between mixing and the horizontal group}
In this section we review the relation between the mixing matrix and the minimal LH symmetry group $\G$ \cite{LAM}.
The relation comes from the recognition that $\G$ is the minimal group generated by a set of 
{\it residual symmetry operators} that
can be computed from the mixing matrix. Conversely, given $\G$, the mixing matrix can be retrieved if certain
computable
vacuum alignments (direction of Higgs expectation values) are used. The rest of this section contains
a sketch of the details to establish terminologies and notations. It also explains how
 the familiar groups $S_4, A_4$, and $S_3$ for leptons arise in this context. 

Let $M_i\ (i=u,d,e,\nu)$ be the Dirac mass matrices for the up-quarks, down-quarks, charged-leptons, and neutrinos
respectively, connecting LH to RH fermions, 
and $M_N$ be the Majorana mass matrix for the heavy (`right-handed') neutrinos. Since we only
deal with LH fermions, the mass matrices we need to consider
 are the LH to LH effective mass matrices $\bar M_i$,
defined by $\bar M_i=M_iM_i^\dagger\ (i=u,d,e)$ and $\bar M_\nu=M_\nu M_N^{-1}M_\nu^T$. Let $U_i\ (i=u,d,e,\nu)$
be the unitary matrices that render $U_i^\dagger\bar M_iU_i\ (i=u,d,e)$ and $U_\nu^T\bar M_\nu U_\nu$ diagonal.
Then the mixing matrix  is $U_{CKM}=U_u^\dagger U_d$ for the quarks and $U_{PMNS}=U_e^\dagger U_\nu$ for 
the leptons. In particular, in the basis where $\bar M_d$ and $\bar M_e$ are diagonal, which we shall adopt
from now on, $U_{CKM}^\dagger=U_u$ and $U_{PMNS}=U_\nu$. In most of the following
 we deal explicitly with the lepton sector, but the 
quark sector can be treated analogously.

Let $F$ and $G$ be respectively unitary symmetry operators for charged leptons and neutrinos, meaning that
$F^\dagger \bar M_e F=\bar M_e$ and $G^T \bar M_\nu G=\bar M_\nu$. Then it follows that $F$ and $\bar M_e$ have 
common eigenvectors, given by the columns of $U_e={\bf 1}$, and $G$ and $\bar M_\nu$ have common
(pseudo-)eigenvectors, given by the columns of $U_\nu=U_{PMNS}$. Being unitary, the eigenvalues of both $F$ and $G$
are unimodular, but on account of the Majorana character of the neutrinos, the eigenvalues of $G$ must also be real,
and by convention one of them is chosen to be $+1$ and two of them chosen to be $-1$. With this convention, three 
symmetry
operators $G_a\ (a=1,2,3)$ can be constructed for each mixing matrix $U_\nu$, in which the eigenvector with 
$+1$ eigenvalue is taken from the $a$th column of $U_\nu$. These operators satisfy $G_a^2=1$ and $G_aG_b=G_c$
if $a,b,c$ are different, so only two of the three are independent.
In contrast, $F$ remains uncertain because it may
have arbitrary eigenvalues, as long as they are unimodular. 

Abstractly, the group generated by $G_a$ is always $Z_2\times Z_2$, whatever the mixing matrix is, and the group
generated by $F$ is always a subgroup of $U(1)$. Obvious that information is useless \cite{LAM, GLL}
if we do not know how this subgroup in the charged-lepton sector
is intertwined with the group $Z_2\times Z_2$ in the neutrino sector. To be useful we need the detailed form
of $F$ and $G_a$ determined from  the mixing matrix to establish the horizontal LH symmetry group $\G$.

The symmetry operators $F$ and $G$ shall be referred
to as {\it residual symmetry operators}, and we will use the notation $\R=(F,G)$ to denote them.
 We know that $F\not=G$ or else there would be no mixing. The minimal horizontal LH symmetry group $\G$
is then the minimal group generated by the operators $F$ and $G$: $\G=\{F,G\}$.

In principle there could be many residual operators of type $F$, but to find 
minimal groups we may confine to only one of them. If we demand $\G$ to be a finite group,
we must have $F^n=1$ for some finite $n$. If we want to be able to 
reconstruct the mixing matrix from the residual symmetry 
operators, we must require $F$ to be non-degenerate, in the sense that no two eigenvalues are identical,
hence $n\ge 3$. Even with
these restrictions we still have an infinite choice of $F$, by picking any $n\ge 3$, 
any three distinct $n$th root of unity as its eigenvalues, and one of the $3!=6$ permuted 
arrangements of eigenvalues in the matrix $F$. Correspondingly, there may be many minimal LH symmetry groups $\G$.

We shall return to the choice of $F$ after discussing the experimental implications on $G_a$. In the neutrino
sector, we shall first assume $U_{PMNS}=U_\nu$ to be of the tri-bimaximal (TBM) form
\be U_\nu={1\over\sqrt{6}}\pmatrix{2&\sqrt{2}&0\cr -1&\sqrt{2}&\sqrt{3}\cr -1&\sqrt{2}&-\sqrt{3}\cr}\label{tribi},\ee
which when compared with experimental data is accurate to within one standard deviation. 
Using the recipe explained above, we can compute from that the three residual symmetry operators to be
\be
G_1&=&{1\over 3}\pmatrix{1&-2&-2\cr -2&-2&1\cr -2&1&-2\cr},\quad
G_2=-{1\over 3}\pmatrix{1&-2&-2\cr -2&1&-2\cr -2&-2&1\cr},\quad
G_3=-\pmatrix{1&0&0\cr 0&0&1\cr 0&1&0\cr}.\label{G}\ee

Now return to the choice of $F$. For $n=3$, the three distinct eigenvalues are the cubic roots of identity,
$1, \o=e^{2\pi i/3}$, and $\o^2$. There are six distinct $F$'s corresponding to the $3!$ permutations, but
only two of them generate distinct groups $\G$. The one generated by $F_1=\diag(1,\o,\o^2)$ is $\G=\{F_1,G_2,G_3\}
=S_4$, the
permutation group of four objects, and the one generated by $F_2=\diag(\o,1,\o^2)$ is 
$\G=\{F_2,G_2,G_3\}=3S_4$, consisting
of $S_4,\o S_4$, and $\o^2 S_4$. It contains $S_4$ as a subgroup. One representation of
$F_1,G_1,G_2,G_3$ in $S_4$ is $F_1=(123),\ G_1=(14),\ G_2=(14)(23),\ G_3=(23)$. This representation
will be useful in seeing how subgroups of $S_4$ are generated.

For any choice of $F$ with $n>3$, it can be shown
that if $\G=\{F,G_2,G_3\}$ is a finite group, then it always contains $S_4$ as a subgroup. Hence $S_4$ is the minimal symmetry group for TBM. 
For example, $PSL_2(7)=\Sigma(168)$ is one of these groups \cite{KL} containing $S_4$ as a subgroup.

The tri-bimaximal matrix implies a vanishing reactor angle $\theta_{13}$, with an ineffective $CP$ phase. 
Since the experimental error of this angle is still large, it is reasonable not to be so committed at the present, which
means that the bimaximal character appeared in the third column of \eq{tribi} should be abandoned. We may still
assume the second column (trimaximal mixing) of \eq{tribi} to be valid \cite{BHS}, 
in which case the first and third columns
of $U_\nu$ are parameterized by two real parameters. With that, the only known symmetry operator 
in the neutrino sector is $G_2$, and
the horizontal group generated by $\{F_1,G_2\}$ can be seen to be $\G=A_4$, the subgroup of $S_4$ with even
permutations. If the two parameters in the mixing matrix are adjusted so that $\theta_{13}=0$, then an accidental
symmetry exists to enlarge the symmetry from $A_4$ to $S_4$. The LH symmetry group generated by other $F$'s
and $G_2$ are not presently known.

Similarly, if we retain the first column of \eq{tribi}, then $\G=\{F_1,G_1\}=S_4$, but in this case the
reactor angle does not necessarily vanish. This illustrates the fact the symmetry alone cannot determine the mixing
matrix. What is needed in addition is the alignment of the composite Higgs, to be discussed in the next section.

If for some reason we only want to retain the third column of \eq{tribi}, namely, the bimaximal mixing, then 
it is well known that the reactor angle vanishes and the atmospheric angle is maximal. In that case the solar
angle is a parameter, and the symmetry group is $\G=\{F_1,G_3\}=S_3$.

\section{vacuum alignment}
Given a LH symmetry group $\G$, the original mixing can be recovered by keeping the
residual symmetry $\R$ intact in the spontaneous breaking of $\G$.
This in turn can be achieved by having the correct vacuum alignments computed as follows. 

We start with an effective Hamiltonian $\bar H$, constructed from LH fermions and composite Higgs fields.
It can be thought of as the result of 
a dynamical Hamiltonian $H$ with the RH fermions integrated out. The Higgs field in $\bar H$ are often
bilinear in the Higgs fields in $H$, hence the name `composite'. 
$\bar H$ may not be local, but it is invariant under the LH symmetry group $\G$.
We retain in $\bar H$ only the Yukawa terms, because only those contribute to the 
mass matrices. Mass terms may be present but they are special cases of Yukawa terms coupled to a singlet Higgs possessing non-zero expectation
values, so they will not be considered separately. 
The composite Higgs coupled to the LH charged leptons will be denoted by $\Phi$, and those coupled to
the LH neutrinos will be denoted by $\Psi$. See Sec.~4 for details.

The condition for $\R=(F,G)$ to remain a symmetry of $\bar H$ after the Higgs acquire their expectation
values is
\be
F^{(A)}\bk{\Phi^{(A)}}&=&\bk{\Phi^{(A)}},\nn\\
G^{(A)}\bk{\Psi^{(A)}}&=&\bk{\Psi^{(A)}},\label{align}\label{FG}\ee
for every residual symmetry operator $G$ and for every IR $(A)$ of the group $\G$. $F^{(A)}$ and $G^{(A)}$
are the IRs of $F$ and $G$.  

For $\G=\{F_1,G_2,G_3\}=S_4$, these alignments have been computed, and the result is listed in Table 1. 
The other three cases in Table 1 are computed similarly using \eq{FG} and the IRs
 of $F^{(A)}$ and $G^{(A)}$, also listed in Table 1. 
Note that the 3-dimensional representation of $S_3$ is always reducible.
In that table, $\G=\{F_1,G_1\}=S_4$ is denoted $\bar S_4$, to distinguish it from $\G=\{F_1,G_2,G_3\}$ which
is denoted as $S_4$. It has a different residual symmetry $\R$, so its Higgs alignments are
not the same as $S_4$. 

In Table 1, the IRs are listed in the second row, with numerals indicating the
dimension of the representation, and $-$ indicating the absence of a particular IR in that  group.
For easiness of printing,
column vectors are written as row vectors in the table, and $(1,1,1)_\perp$ is any vector
orthogonal to $(1,1,1)$. The overall normalization of each vacuum alignment is arbitrary because \eq{FG} is linear.
`dr' in the first row stands for `defining representation', which is the representation
the LH fermions belong to. I apologize for the convention adopted here in which the defining representation of $S_4$,
for which $\det(G_i)=+1$, is called $\dc'$ rather than $\dc$. The other symbols are,
$\o=e^{2\pi i/3}$, $\sigma_1$ is the Pauli matrix $[[0,1],[1,0]]$,  and $g_a$
are the $G_a$ defined in eq.~\eq{G}, namely, $g_1=[[1,-2,-2],[-2,-2,1],[-2,1,-2]]/3$,
$g_2=-[[1,-2,-2],[ -2,1,-2],[ -2,-2,1]]/3$, and
$g_3=-[[1,0,0],[ 0,0,1],[ 0,1,0]]$. Note that $\bk{\Psi}$ in the $\dc'$ IR is (0,0,0) for $S_4$ but
$(1,1,1)_\perp$ for $\bar S_4$. It is this difference that in principle allows $\bar S_4$ 
to accommodate non-zero $\theta_{13}$, though the discussion in the next section shows that this turns
out to be impossible.

\vspace{.5cm}

\begin{center}
\begin{tabular}{|l|r|c|c|c|c|c|c|}
\hline
dr&&$S_3$&$-$&$-$&$S_3$&$A_4$&$S_4,\bar S_4$\\ \hline\hline
&&\da&$\da'$&$\da''$&\db&\dc&$\dc'$\\ \hline\hline
$S_4$&$F_1$&1&$1$&$-$&$\diag(\o,\o^2)$&$\diag(1,\o,\o^2)$&$\diag(1,\o,\o^2)$\\ \cline{2-8}
&$G_2$&1&$1$&$-$&$\diag(1,1)$&$g_2$&$g_2$\\ \cline{2-8}
&$G_3$&1&$$-1$$&$-$&$\sigma_1$&$-g_3$&$g_3$\\ \cline{2-8}
&$\bk{\Phi}$&1&1&$-$&$(0,0)$&$(1,0,0)$&$(1,0,0)$\\ \cline{2-8}
&$\bk{\Psi}$&1&0&$-$&$(1,1)$&$(1,1,1)$&$(0,0,0)$\\ \hline\hline
$A_4$&$F_1$&1&$\o$&$\o^2$&$-$&$\diag(1,\o,\o^2)$&$-$\\ \cline{2-8}
&$G_2$&1&1&1&$-$&$g_2$&$-$\\ \cline{2-8}
&$\bk{\Phi}$&1&0&0&$-$&$(1,0,0)$&$-$\\ \cline{2-8}
&$\bk{\Psi}$&1&1&1&$-$&$(1,1,1)$&$-$\\ \hline\hline
$S_3$&$F_1$&1&1&$-$&$\diag(\o,\o^2)$&$-$&$-$\\ \cline{2-8}
&$G_3$&1&$-1$&$-$&$\sigma_1$&$-$&$-$\\ \cline{2-8}
&$\bk{\Phi}$&1&1&$-$&$(0,0)$&$-$&$-$\\ \cline{2-8}
&$\bk{\Psi}$&1&0&$-$&$(1,1)$&$-$&$-$\\ \hline\hline
$\bar S_4$&$F_1$&1&$1$&$-$&$\diag(\o,\o^2)$&$\diag(1,\o,\o^2)$&$\diag(1,\o,\o^2)$\\ \cline{2-8}
&$G_1$&1&$$-1$$&$-$&$\sigma_1$&$-g_1$&$g_1$\\ \cline{2-8}
&$\bk{\Phi}$&1&1&$-$&$(0,0)$&$(1,0,0)$&$(1,0,0)$\\ \cline{2-8}
&$\bk{\Psi}$&1&0&$-$&$(1,1)$&$(1,1,1)$&$(1,1,1)_\perp$\\ \hline

\end{tabular}
\end{center}
\vspace{.5cm}
Table 1. Vacuum alignments $\bk{\Phi}$ and $\bk{\Psi}$ of the horizontal symmetry groups
$S_4, A_4, S_3, \bar S_4$ for leptons.  See the text for an explanation of the symbols.

\section{low energy content}
As mentioned in the Introduction, the low-energy content (masses,
mixing angles, and phases) of any dynamical theory is already contained in the LH effective mass matrices
$\bar M_i\ (i=u,d,e,\nu)$, which can be read off from the LH effective Hamiltonian $\bar H$. 
In this section we shall discuss the structure of $\bar H$ for every symmetry
group $\G$ discussed in the previous section.

The effective Hamiltonian can be written symbolically as
\be
\bar H=\sum_A\(\lambda_A e_L^\dagger e_L\Phi^A+\mu_A \nu_L^T \nu_L\Psi^A\),\label{heff}\ee
with all couplings  understood to be $\G$-invariant. 
The LH charged-lepton fields are denoted by
$e_L$, and the LH neutrino fields are denoted by $\nu_L$. 
They belong to the defining representations ${\bf r}=\dc',\dc,(\da,\db),\dc'$,
respectively for the groups $\G=S_4,A_4,S_3,\bar S_4$ (see Table 1). 
Only IRs $A$ contained in the Clebsch-Gordan (CG) series
${\bf r}\times{\bf r}$ appear in the sum in \eq{heff}.
The effective mass matrices $\bar M_e$ and 
$\bar M_\nu$ can be read off from \eq{heff} using the Higgs alignments  listed in Table 1. 
Since $\bar M_e=\bar M_e^\dagger$ and $\bar M_\nu=\bar M_\nu^T$, IR $A$ that leads to an anti-symmetric 
contribution to $\bar M_\nu$ should be dropped. Moreover, the parameters $\l_A$ have to be real though
$\mu_A$ may be complex. 

The construction of \eq{heff} is  schematically shown in Table 2. 
The first and second rows together specify the defining representation ${\bf r}$ of the various groups, and 
the remaining rows indicate how the CG series ${\bf r}\times{\bf r}$ pans out.
$-$ denotes an IR that is not present in that group, and \x\
denotes an IR that does not appear in the CG series ${\bf r}\times{\bf r}$. A  `0' indicates a term
in which the expectation value $\bk{\Phi}$ or $\bk{\Psi}$ vanishes so the corresponding coupling constant
does not appear as a parameter. Each entry in $S_3$ is shown as a $2\times 2$ matrix, indicating the Clebsch-Gordan series for
$\pmatrix{$\da\x\da$&$\da\x\db$\cr $\db\x\da$&$\db\x\db$\cr}$.

\vspace{.5cm}

The directions of the Higgs expectation values are given in Table 1, and 
their magnitudes can be absorbed into the Yukawa coupling constants $\l_A$ or $\mu_A$, hence the number of free
parameters is just the number of Yukawa coupling constants, each of which is represented by a black dot,
with the total number listed in the last column.
An open circle indicates an antisymmetric contribution
to $\bar M_\nu$ that has been dropped. In the case of $S_3$, the open circle actually represents the antisymmetric
combination of $\da\times\db$ and $\db\times\da$.
As mentioned above, the parameters in the $\Phi$ rows have to be real
but the parameters in the $\Psi$ rows may be complex.

\vspace{1cm}

\begin{center}
\begin{tabular}{|l|r|c|c|c|c|c|c|c|}
\hline
&${\bf r}$&$S_3$&$-$&$-$&$S_3$&$A_4$&$S_4,\bar S_4$&\\ \hline\hline
&$A$&\da&$\da'$&$\da''$&\db&\dc&$\dc'$&\#\\ \hline\hline
$S_4$&$\Phi$&$\bullet$&\x&$-$&$0$&\bl&\bl&3\\ \cline{2-9}
&$\Psi$&\bl&\x&$-$&\bl&\bl&0&3\\ \hline\hline
$A_4$&$\Phi$&\bl&0&0&$-$&\bl\ \bl&$-$&3\\ \cline{2-9}
&$\Psi$&\bl&\bl&\bl&$-$&\bl\ $\circ$&$-$&4\\ \hline\hline
$S_3$&$\Phi$&$\matrix{\bullet&\times\cr \times&\bullet\cr}$&$\matrix{\times&\times\cr\times&\bullet\cr}$&$-$&
$\matrix{\times&0\cr 0&0\cr}$&$-$&$-$&3\\ \cline{2-9}
&$\Psi$&$\matrix{\bullet&\times\cr \times&\bullet\cr}$&$\matrix{\times&\times\cr\times&0\cr}$&$-$&$\matrix{\times&\bullet\cr\circ&\bullet}$&$-$&$-$&4\\ \hline\hline
$\bar S_4$&$\Phi$&$\bullet$&\x&$-$&$0$&\bl&\bl&3\\ \cline{2-9}
&$\Psi$&\bl&\x&$-$&\bl&\bl&$\circ$\ &3\\ \hline\hline
\end{tabular}
\end{center}

\noindent Table 2. Coupling scheme for $\bar H$ in \eq{heff}, with the total number of independent free parameters listed
in the last column. See the text before the table for an explanation of the symbols.

\vspace{.5cm}

We want to find out whether  mass matrices so obtained contain less, just enough, or more parameters than
 necessary to fit the low-energy data. In  the charged-lepton sector,
every $\Phi$ row of Table 2 contains exactly
three real parameters to fill the three entries of the diagonal mass matrix $\bar M_e$, 
which is just enough to fit the three charged-lepton masses no matter what $\G$ is.

The neutrino sector is more complicated. The three neutrino masses combining with the two Majorana phases form 
three complex masses, albeit with an arbitrary overall phase. As to mixing, since $S_4$
 automatically gives rise to the tri-bimaximal matrix, no additional parameter is required to fix the mixing,
so the total number of complex parameters needed to fit the low energy data is 3, exactly what the 
effective Hamiltonian provides. 
In the other three cases, where only one column of the tri-bimaximal matrix is
automatically obtained, we need one additional complex parameter to fully specify the mixing matrix \cite{TEXTURE}. 
Thus in those three
cases the total number of complex parameters needed to fit the low energy data is 4. From Table 2, we see that
is precisely what we have for $A_4$ and $S_3$, but we are one short in $\bar S_4$. As explained below, it turns
out in this case that $\theta_{13}=0$ automatically, so all the mixings are determined, and 3 is just the right number 
of parameters to fit the complex neutrino masses.

Let us now look at the resulting
neutrino mass matrices to understand what really happens. 
For $A_4$, explicit calculation shows that
\be
\bar M_\nu=\pmatrix{\bar\mu_1+2\bar\mu_3&\bar\mu_{1''}-\bar\mu_3&\bar\mu_{1'}-\bar\mu_3\cr
\bar\mu_{1''}-\bar\mu_3&\bar\mu_{1'}+2\bar\mu_3&\bar\mu_1-\bar\mu_3\cr
\bar\mu_{1'}-\bar\mu_3&\bar\mu_1-\bar\mu_3&\bar\mu_{1''}+2\bar\mu_3\cr}\ee
where $\bar\mu_1=\mu_1/\sqrt{3},\ \bar\mu_{1'}=\mu_{1'}/\sqrt{3},\ \bar\mu_{1''}=\mu_{1''}/\sqrt{3},\ 
\bar\mu_3=\mu_3/\sqrt{6}$. This matrix is magic \cite{MAGIC}, 
whose row and column sums  all equal to $\bar\mu_1+\bar\mu_{1'}+
\bar\mu_{1''}$, hence the mixing matrix computed from it is trimaximal. It has four complex parameters, which can
be used to fit the three complex neutrino masses and the remaining mixing parameter. In particular, if 
$\bar\mu_{1'}=\bar\mu_{1''}\doteq \bar\mu_2$, then the mass matrix is also 2-3 symmetric \cite{23}, and the corresponding
mixing is bimaximal. The mass matrix in that case coincides with the mass matrix of $S_4$, with the $\db$ IR
of $S_4$ decomposing into $(\da',\da'')$ in $A_4$, and the corresponding mixing matrix tri-bimaximal.
In summary, the $A_4$ effective Hamiltonian $\bar H$ contains one more complex parameter than $S_4$, and that can be used to accommodate a
non-vanishing reactor angle $\theta_{13}$ and a CP phase. If that parameter is chosen so that the reactor angle vanishes,
then the $A_4$ symmetry is accidentally enlarged to an $S_4$ symmetry.

The neutrino mass matrix for $S_3$ is
\be
\bar M_\nu=\pmatrix{\mu_1&\mu_2&\mu_2\cr \mu_2&\bar\mu_2&\bar\mu_1\cr \mu_2&\bar\mu_1&\bar\mu_2\cr},\ee
where $\mu_1$ is the Yukawa coupling of $\da\times\da$ to $A=1$, and $\bar\mu_1$ is the Yukawa coupling of 
$\db\times\db$ to $A=1$ divided by $\sqrt{2}$. Similarly, $\mu_2$ and $\bar\mu_2$ are the coupling constants
for $A=2$ coming from $\da\times\db$ and $\db\times\db$ respectively. This mass matrix is 2-3 symmetric,
befitting a bimaximal mixing appropriate to $S_3$. It has four adjustable complex parameters that can
be used to fit the three complex neutrino masses and the remaining complex mixing parameter. In particular,
if $\mu_1+\mu_2=\bar\mu_1+\bar\mu_2$, the mass matrix becomes magic, the mixing becomes tri-bimaximal,
and $S_3$ attains the enlarged accidental symmetry $S_4$.

Finally let us consider $\bar S_4$. It differs from $S_4$ in that $\bk{\Psi}$ for $\dc'$ does not have to
vanish (see the last row of Table 1).
 However, since the $\dc'\times\dc'\to\dc'$ coupling is antisymmetric on the left, it cannot
contribute to $\bar M_\nu$, so the mass matrices in $\bar S_4$ are identical to the mass matrices in $S_4$.
In other words, it automatically predicts a zero reactor angle $\theta_{13}$, which is why only 3 complex
parameters are enough to fit all the low energy data.

By using less parameters than those listed in the last column of Table 2, one can construct models
with `predictions' relating some of the low-energy quantities, {\it e.g.,} mixing angles in terms of mass ratios.
It should also be noted that local dynamical models may contain more parameters than those listed
in Table 2, but if that is the case, some of them must combine in their corresponding LH effective Hamiltonian.

In conclusion, if tri-bimaximal neutrino mixing is exactly true, then $S_4$ and $\bar S_4$ are the most
economical symmetry group because it makes the most predications and requires the smallest number of parameters
to fit the remaining data. On the other hand, if $\theta_{13}\not=0$, then only $A_4$ among these four 
could be the correct symmetry group, assuming of course that trimaximal mixing still remains valid with
better data. If neither trimaximal nor bimaximal mixing is valid in the face of better data, 
then these four groups can
only lead to an approximate fitting in tree order. Which of the four gives a better approximation then
depends on the data and the results of higher-order calculations.

\section{quark mixing}
Quark mixing is very different from neutrino mixing because it is small. In principle
the method reviewed in Sec.~2 can be used to find its horizontal symmetry group $\G$, but in practice,
it is difficult to make $\G$  a finite group because there is no known parameterization in this sector equivalent to
the tri-bimaximal matrix. 
For $\G$ to be a finite group, there must be an integer $n$ so that $g^n=1$ for every $g\in G$.
If this property is true for some matrix $g$, it can no longer be true by suitably altering the matrix elements
a little bit within their experimental error.  This is similar to saying that 22/7 is a rational number but
$\pi$ is not, although the former is an approximation to the latter. 
If a finite group exists, it is likely that this group belongs to a series with an adjustable
parameter so that the small Cabibbo angle can be accommodated. These would suggest finite groups like 
the dihedral groups $D_m$, or the $SU(3)$ finite subgroups $\Delta(3m^2)$ and
$\Delta(6m^2)$. Indeed $D_7$ and $D_{14}$ have been proposed as candidates \cite{LAM,BHH}.

However, if we want quarks and leptons to be united, like in a grand unified theory, then they
 should have the same horizontal symmetry group $\G$. There are two ways to accommodate that. One is to find
a large group containing both, such as $SU(3)$, which at an early stage breaks down to the individual symmetry groups
in the quark and the lepton sectors. Such a construction would have to be completely dynamical and
the present approach can say very little about it. The other is to assume the leptonic symmetry group,
like $S_4, A_4$, or $S_3$, to be  also the symmetry group of the quarks, with the same set of Higgs and the
same vacuum alignments. Many dynamical models have been constructed along these lines. In that case,
quarks do not mix in the tree level, so all their mixing must come from loop corrections. One way to do that
is to assign both the LH and the RH fermions to the defining 3-dimensional representation, and introduce two
sets of Higgs, $\phi$ and $\psi$, so that $\phi$ have the same vacuum alignments as $\Phi$, and $\psi$
has the same vacuum alignments as $\Psi$. The Higgs $\f$
would couple the LH
fermions to the RH fermions to produce the diagonal Dirac mass matrices $M_i\ (i=u,d,e,\nu)$, and $\psi$ would couple only to the heavy Majorana neutrinos
to produce the mass matrix $M_N$. This would guarantee 
$\bar M_i=M_i^\dagger M_i$ to be diagonal, and the neutrino effective mass matrix $\bar M_\nu=M_\nu^TM_N^{-1}M_\nu$
to give rise to a neutrino mixing appropriate to $\G$. How realistic schemes like that are depend on how successful
are the high-order corrections. These corrections can be implemented by higher-dimensional terms, at the expense of
introducing new parameters, but it would be nice if they could be computed in a renormalizable theory.

I thank Christoph Luhn for helpful discussions and comments.

\end{document}